# Growth of EuO single crystals at reduced temperatures

Daniel C. Ramirez,[1,2] Tiglet Besara,[1] Jeffrey B. Whalen,[1] and Theo Siegrist[1,3,*]

[1] *National High Magnetic Field Laboratory, Florida State University, Tallahassee, FL 32310, USA*
[2] *Materials Science and Engineering, Florida State University, Tallahassee, FL 32310, USA*
[3] *Department of Chemical and Biomedical Engineering, FAMU-FSU College of Engineering, Florida State University, Tallahassee, FL 32310, USA*

**Abstract**

Single crystals of $(Eu_{1-x}Ba_x)O$ have been grown in a molten barium-magnesium metal flux at temperatures up to 1000°C, producing single crystals of $(Eu_{1-x}Ba_x)O$ with barium doping levels ranging from $x=0.03$ to $x=0.25$. Magnetic measurements show that the ferromagnetic Curie temperature $T_C$ correlates with the Ba doping levels, and a modified Heisenberg model was used to describe the stoichiometry dependence of $T_C$. Extrapolation of the results indicates that a sample with Ba concentration of $x=0.72$ should have a $T_C$ of 0 K, potentially producing a quantum phase transition in this material.

**Introduction**

EuO is the first ferromagnetic semiconductor discovered and investigated in details [1, 2]. The simple rock-salt structure of EuO makes this material a model compound for studying magnetic interactions, especially since its ferromagnetic moment per volume exceeds the value for iron. The highly coupled electronic and magnetic interactions in EuO give rise to interesting physical properties that reflect this coupling. For instance, the ferromagnetic order changes the electronic band structure in a way that breaks the degeneracy of the spin up and spin down bands, and shifts the optical band gap [3-5]. In addition, an insulator-metal transition at $T_C$ [6, 7], and a giant magneto-optical Kerr effect have been observed [8]. The interest in EuO has seen a considerable upswing recently due to its potential as a spintronics material. For example, EuO devices were employed to for spin filtering [9-12], the interface between EuO and GdN is expected to produce a band inversion and a nonzero Chern number [13], a giant spin-phonon coupling has been reported [14], it has been suggested that a CdO/EuO superlattice is a Weyl semimetal [15], and the Hall effect of EuO thin films has been observed [16, 17]. Efforts to increase the Curie temperature via doping and alloying of EuO have also seen a resurgency: Curie temperatures up to 134 K in single crystals and even higher in thin films have been achieved [18-27].

EuO has been synthesized as nanostructural materials [28-30], and epitaxial thin films of EuO have been grown for integration in various types of devices [31-34] and more recently EuO has been grown on graphene [35-38]. Prior to this report, single crystal growth of EuO used high temperature techniques exclusively. The general procedure required $Eu_2O_3$ and Eu metal to be sealed in a refractory metal crucible (molybdenum, tantalum or tungsten) and heated to temperatures between 1700 and 2300°C [39-43] to reduce the trivalent $Eu^{3+}$ in $Eu_2O_3$ to the divalent $Eu^{2+}$ in EuO. Single crystals with cubic habit and several millimeters' edge length were obtained in this way. However, apart from thin film growth, no bulk single crystal growth has been reported where growth temperatures are below 1000°C. When using $Eu_2O_3$ as starting material, the reduction of $Eu^{3+}$ to $Eu^{2+}$ is achieved only in a strongly reducing environment. For powder preparation, low temperature synthetic routes employed either EuO$X$ ($X$ = halogen) and LiH [44], or mixtures of $Eu_2O_3$ and $EuH_2$ [45]. Crystal growth of EuO at low temperatures may be achieved using chemical transport methods or a flux. In the case of flux growth, the flux requirements are: low melting temperature, good solubility for $Eu^{2+}$, solubility for oxygen, and potential to reduce $Eu_2O_3$.

Here, we describe a method to grow EuO single crystals at relatively moderate temperatures – below 1000°C – using an alkaline earth metal flux. As $Eu^{2+}$ is chemically similar to the alkaline earth metals, good solubility in Ca, Sr and Ba metal is expected. It has also been found that oxygen has a substantial solubility in alkaline earth metal fluxes [46]. Since full miscibility in the SrO-EuO system has been reported [45], a Sr-based flux is excluded, leaving Ca and Ba as potential fluxes. However, inclusion of the flux is expected, but the ionic size differences between $Eu^{2+}$, $Ca^{2+}$ and $Ba^{2+}$ are likely to limit the solid solution range. The high oxygen solubility of molten barium makes barium the preferred flux, with magnesium added to adjust the reducing power of the flux.

**Experimental Section**

Reaction of $Eu_2O_3$ in a Ba-Mg flux yielded single crystals of $(Eu_{1-x}Ba_x)O$. The solubility of $Eu_2O_3$ in molten Ba has been demonstrated previously [46]. The starting molar ratios of Ba:Mg:$Eu_2O_3$ ranged from 20:2:2 to 20:5:2, and all materials were loaded into stainless steel crucibles welded shut in an argon atmosphere. The steel crucibles were then sealed in evacuated quartz ampoules that were subsequently placed in a muffle furnace, heated to 1000°C at a rate of 10°C/h, soaked for 20 h, and then cooled at a rate of 1.3°C/h to 800°C. The ampoules were then removed from the furnace, inverted and centrifuged to separate the flux from the crystals.

Crystals obtained by this method were stable in ambient environment for short periods of time, but deteriorated due to hydroxide formation within 24 hours when left unprotected. Extracted crystals were therefore stored under argon to avoid contamination by moisture. Under dry conditions, measurements of crystals stored for several months consistently returned the same results.

Single crystal X-ray diffraction was carried out with an Oxford Diffraction Xcalibur2 CCD diffractometer with graphite monochromated MoKα radiation. Energy dispersive spectroscopy (EDS) with a JEOL 5900 scanning electron microscope was used to verify the Ba:Eu ratio, together with mass spectroscopy on selected samples, with consistent results.

Magnetic susceptibility measurements were performed using a Quantum Design SQUID magnetometer in an applied field of 0.1 T. M-H loops of the samples with cube shape were measured at 1.8 K with the magnetic field along the [110] direction (face diagonal).

**Results and Discussion**

The crystal growth of EuO in the barium-magnesium flux occurs on self-nucleated sites, resulting in several cube-shaped crystals of sizes often exceeding 1 mm on edges (Fig. 1). Initial experiments showed that a pure barium flux is capable of reducing $Eu_2O_3$ to EuO, but only a handful of small crystals with irregular shape were obtained in this fashion. To adjust the reducing power of the flux, while simultaneously retaining good solubility for EuO, magnesium metal was added to the barium flux. The magnesium is responsible for the enhanced reduction of $Eu_2O_3$ to EuO, with minor production of MgO forming as transparent MgO crystals embedded throughout the flux according to $Eu_2O_3 + Mg \rightarrow 2EuO + MgO$.

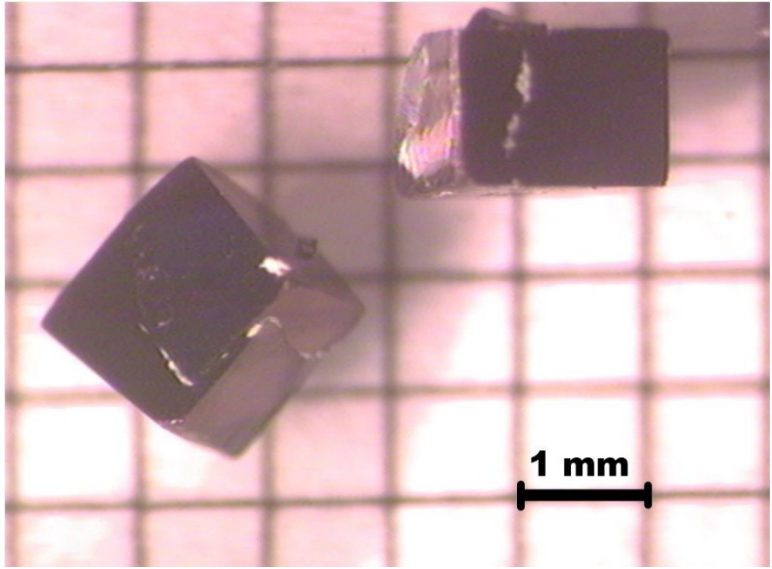

FIG. 1. Single crystals of $Eu_{1-x}Ba_xO$ grown from a Ba-Mg eutectic flux. The crystals are cubic in morphology due to the underlying fcc structure, and show (100) faces.

The ionic size difference between $Ba^{2+}$ and $Eu^{2+}$ is not sufficient to completely eliminate the incorporation of barium into EuO, therefore producing ternary $(Eu_{1-x}Ba_x)O$ single crystals. Since the lattice energies for EuO (–3267 kJ/mol) and BaO (–3054 kJ/mol) differ, the barium incorporation into EuO can be tuned by adjusting the growth temperature, the cooling rate, and the temperature when the ampoules are removed from the furnace. The results from several growth runs demonstrated that the barium content $x$, as determined by EDS and Vegard's law, can be reproducibly adjusted from $x \approx 0.03$ to $x \approx 0.25$. Interestingly, a systematic decrease in barium content is observed when adding magnesium metal to the flux while keeping all other growth parameters constant. We surmise that the addition of Mg enhances the stability of barium in the liquid-phase as the mixture approaches a deep low melting eutectic at Ba:Mg = 13:7 [47], therefore reducing Ba inclusion in the crystals.

Unit cell parameters for the various crystals, as determined by collection of high angle frames on the CCD diffractometer are summarized in Table 1. Eu/Ba ratios were also determined using energy dispersive X-ray analysis (EDS) and obtained via magnetic measurements. All results are consistent with Vegard's law and the expected magnetic properties.

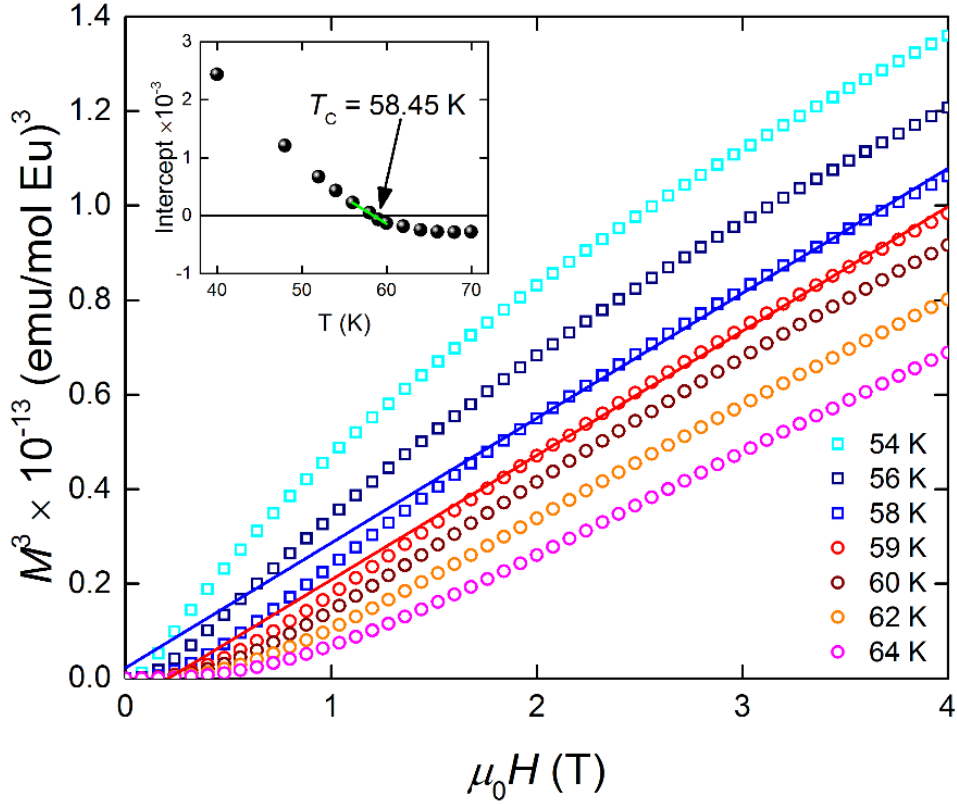

FIG. 2. Arrott isothermal magnetization curves of $M^3$ versus applied field. Data are fit in the linear region at high field strengths (1.75 T to 4 T). The isothermal magnetization curve with an intercept at zero corresponds to the Curie temperature $T_C$. A linear fit of the intercepts is shown in the inset, providing $T_C$ = 58.45 K.

Ferromagnetic behavior was observed in all crystals below 69 K which is the $T_C$ for pristine EuO [1]. In order to obtain a reference point for $T_C$ from the magnetization vs. temperature curves, the Curie temperature of one representative crystal was determined via magnetization isotherms and an Arrott plot, giving a $T_C$ = 58.45 K for a sample with $x \approx 0.088$, as illustrated in Fig. 2. In Fig. 3, the $T_C$ = 58.45 K is indicated in the susceptibility plot obtained in an applied field of 0.1 T, located closely to 1/3 of the saturation magnetization. In this way, an "Arrott criterion" for $T_C$ was established and the $T_C$ of all the crystals could be easily determined from the susceptibility curves. It should be noted that this criterion is established with an applied field // [110] and is therefore expected to be most accurate for samples measured along this orientation. It should also be noted that the actual Curie temperature may differ slightly, as the "Arrott criterion" may deviate somewhat from the value of ≈1/3 for different stoichiometries.

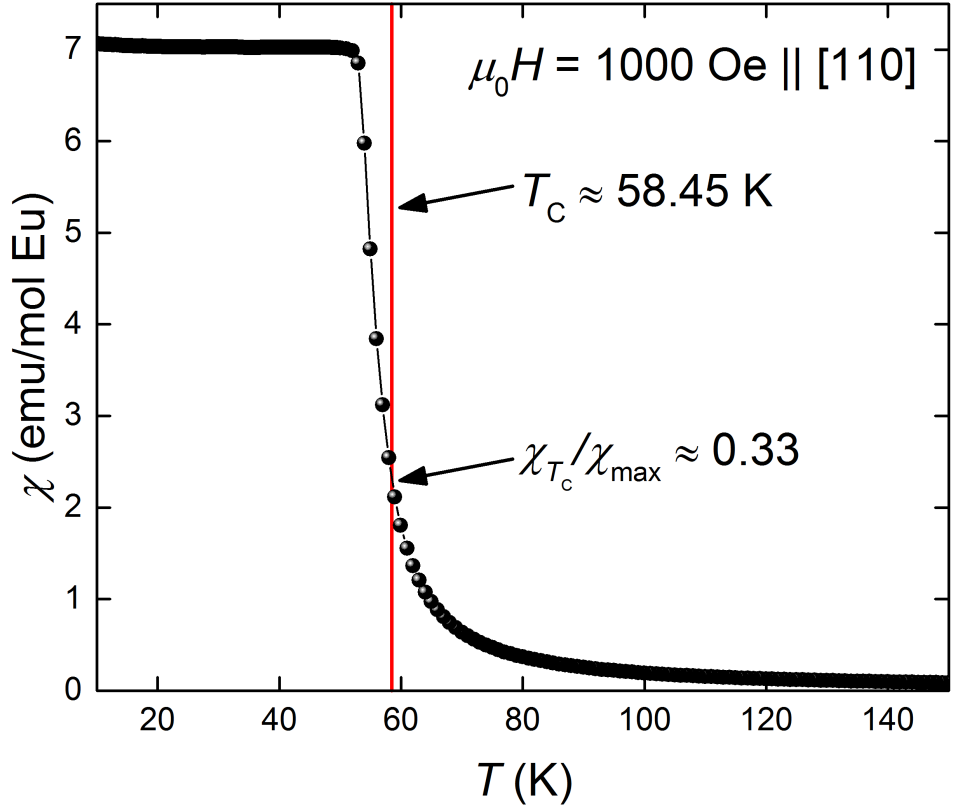

FIG. 3. Susceptibility vs. temperature at an applied field of 1000 Oe along [110] of a crystal of $(Eu_{0.91}Ba_{0.09})O$. The red line indicates the $T_C$ = 58.45 K as given by the Arrott analysis (see Fig. 2). This temperature corresponds to a point on the curve at approximately 1/3 of the saturation magnetization.

A modified two-parameter Curie-Weiss fit of the temperature-dependent susceptibility ($\chi$) measurements was used to determine the Weiss field parameter $\theta$ while simultaneously estimating the Ba content $x$ of each sample. This is represented by

$$\chi = \frac{m}{M_{FU}} \cdot \frac{(1-x)N_A \mu_{eff}^2 \mu_B^2}{3k_B(T-\theta)}, \quad (1)$$

where $m$ denotes sample mass and

$$M_{FU} = x \cdot M_{BaO} + (1-x) \cdot M_{EuO} \quad (2)$$

denotes formula mass of each sample as a function of $x$. This expression is then used to fit the $\chi$ vs. $T$ data, treating $x$ and $\theta$ as fit parameters, while keeping $\mu_{eff}$, taken as the free ion moment of 7.94 $\mu_B$ per $Eu^{2+}$ ion, fixed. The results are listed in Table 1.

Table 1. Summary of structural and magnetic results for $Eu_{1-x}Ba_xO$ crystals, listing the unit cell parameter, barium content $x$ from Vegard's law and from a Curie-Weiss fit, and Curie ("Arrott criterion" $T_C$) and Weiss ($\theta$) temperatures.

| Lattice parameter (Å) | $x$ Vegard's law | $x$ Curie-Weiss | Arrott criterion $T_C$ (K) | Weiss temperature $\theta$ (K) |
|---|---|---|---|---|
| 5.2313 | 0.222(1) | 0.228(3) | 49.3(1) | 50.6(1) |
| 5.2175 | 0.187(1) | 0.184(3) | 48.5(1) | 50.2(1) |
| 5.2027 | 0.149(1) | 0.156(4) | 53.5(1) | 54.0(1) |
| 5.2003 | 0.144(3) | 0.150(7) | 53.4(1) | 55.4(1) |
| 5.1902 | 0.118(1) | 0.124(2) | 57.5(1) | 58.4(1) |
| 5.1871 | 0.110(2) | 0.117(5) | 55.8(1) | 57.2(1) |
| 5.1868 | 0.109(2) | 0.116(1) | 56.5(1) | 57.1(1) |
| 5.1728 | 0.074(5) | 0.074(1) | 58.5(1) | 58.8(1) |
| 5.1682 | 0.062(2) | 0.070(1) | 61.3(1) | 62.1(1) |
| 5.1640 | 0.052(2) | 0.063(2) | 65.2(1) | 64.3(1) |
| 5.1632 | 0.049(1) | 0.048(1) | 61.3(1) | 65.1(1) |
| 5.1556 | 0.031(2) | 0.038(2) | 64.8(1) | 64.8(1) |

Comparing $\theta$ and the "Arrott criterion" $T_C$ shows that the two values for each sample are within a few degrees of each other, providing a consistent way to determine both $\theta$ and $T_C$. The influence of the Ba substitution in $Eu_{1-x}Ba_xO$ on $T_C$ can therefore be studied. As shown in Fig. 4 and Table 1, Curie temperatures $T_C$ range from 50 K for $x=0.22$ to 65 K for $x=0.03$, compared to a $T_C = 69$ K for pristine EuO [1]. With the Curie temperature defined via the "Arrott criterion", and the Weiss constant and Ba content $x$ determined from temperature-dependent susceptibility measurements, the effect of the barium substitution on the magnetic interaction strength can now be determined.

The magnetic interactions in EuO are mediated by the nearest neighbor (NN) indirect exchange mechanism $J_1$ and the next-nearest neighbor (NNN) superexchange mechanism $J_2$. The 12 NN interactions consist of a virtual transition of a $4f$ electron to an empty $5d$ state, while the 6 NNN interactions rely on $s$-$f$ coupling within the Eu atom, and $s$-$p$ interactions between europium and oxygen [48, 49].

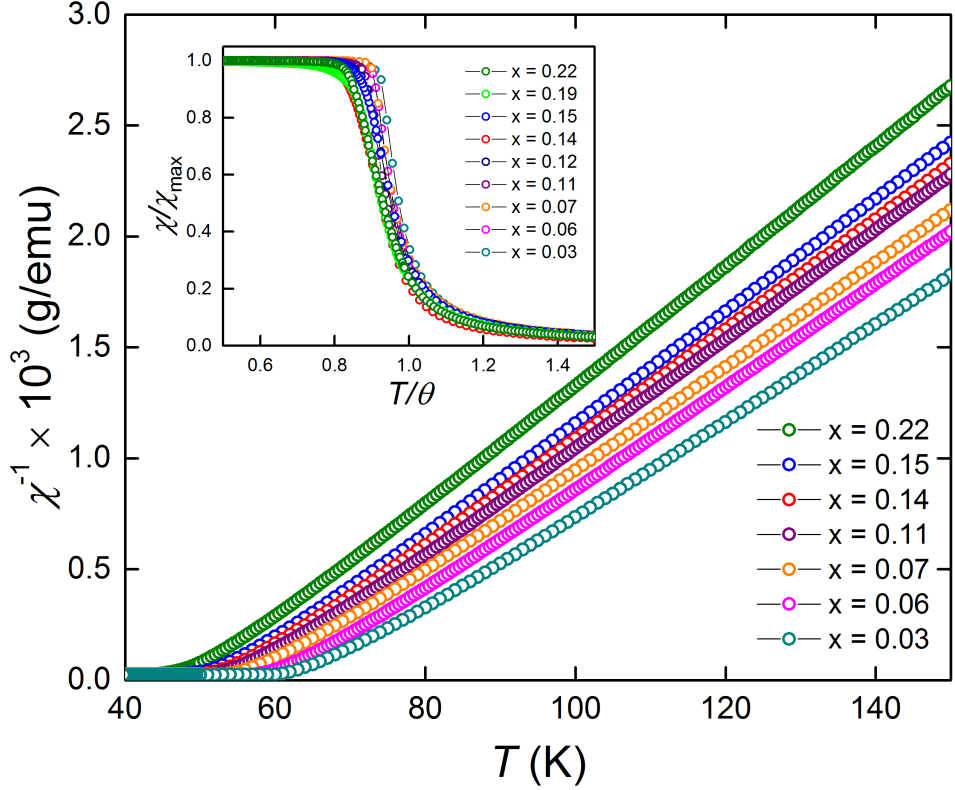

FIG. 4. Inverse susceptibility vs. temperature for selected Ba contents. Inset: Eu$_{1-x}$Ba$_x$O reduced susceptibility $\chi/\chi_{max}$ vs. reduced temperature $T/\theta$ for $x = 0.03$ to $0.22$.

While it is clear that $J_1$ is ferromagnetic, whether $J_2$ is ferromagnetic or antiferromagnetic is still a matter of debate (see Refs. [50] and [51] and references therein). These exchange interactions depend strongly on the Eu-Eu distance, and in the isostructural chalcogenide series, this is responsible for the transition to antiferromagnetism in rock salt-type EuTe, with a lattice parameter of 6.591 Å [48]. However, the absolute value of $J_2$ is small compared to $J_1$. The incorporation of Ba into the EuO lattice in our case expands the cubic lattice parameter by less than 2%, so that the changes in the ferromagnetic NN and NNN interactions are expected to scale linearly with the barium content $x$. This assumption can be further justified when the susceptibility measurements are scaled by using a reduced susceptibility $\chi/\chi_{max}$ (where $\chi_{max}$ is taken at 20 K) and a reduced temperature $T/\theta$ (or $T/T_C$). In the inset to Fig. 4, this reduced susceptibility is plotted against the reduced temperature, showing that all measured crystals follow essentially the same behavior, as expected. Deviations from the universal curve are due to small variations in the sample alignment in the magnetometer and varying demagnetization factors due to differences in the sample shape.

EuO and the other europium monochalcogenides are well known Heisenberg ferromagnets; therefore, the $T_C$ can be estimated using the mean-field model

$$T_C = \frac{2}{3k_B} S(S+1)(Z_1 J_1 + Z_2 J_2), \qquad (3)$$

where $Z_1=12$ is the number of NNs, $Z_2=6$ is the number of NNNs, and $S=7/2$. This model can be modified to include the effects of barium substitution. Two effects must be considered: the reduction of the ferromagnetic interactions due to the substitution of non-magnetic Ba atoms on Eu sites, and the dependence of the exchange interaction on the lattice parameter induced by the larger Ba atoms, i.e., $T_C$ will depend on the barium content $x$ and the lattice parameter $a$. In addition, we assume a linear dependence of the lattice parameter on the barium content, an assumption that is justified given that Vegard's law is followed (columns 1 and 2 in Table 1). The linear fit of the plot (see inset to Fig. 5) yields

$$a(x) \approx 0.395x + 5.144 = \gamma x + a_0, \qquad (4)$$

where $a_0$ is the lattice parameter of pristine EuO ($x=0$). Since the absolute value of $J_2$ is small compared to $J_1$, we will combine both interactions in a single term in our model, an effective $J_{\text{eff}}$. However, $T_C$ will be affected both by the reduction of the interactions due to barium inclusion and by the increase of the lattice parameter, therefore prompting a dependence on $x$ and $a$, viz. $J_{\text{eff}} = J_{\text{eff}}(x,a)$. Eqn. (3) can then be rewritten as

$$T_C(x,a) = \xi\big(12 J_1(x,a) + 6 J_2(x,a)\big) \approx \xi J_{\text{eff}}(x, a(x)), \qquad (3')$$

where we have introduced the shorthand notation $\xi \equiv \frac{2}{3k_B} S(S+1)$. The influence of the lattice expansion and the barium content on the ferromagnetic transition can now be modeled by a Taylor expansion of the model Eqn. (3') around $x = 0$ and $a = a_0$:

$$T_C(x,a) \approx \xi J_{\text{eff}}(0, a_0) + 2\xi\gamma \frac{\partial J_{\text{eff}}(0, a_0)}{\partial a} x + 2\xi\gamma^2 \frac{\partial^2 J_{\text{eff}}(0, a_0)}{\partial a^2} x^2, \qquad (5)$$

where we have utilized the linear expression in Eqn. (4) and the chain rule to express the interaction terms in derivatives of $a$, and $\gamma$ is the slope of the linear relation between $a$ and $x$ derived from Vegard's law. The first term (zeroth order coefficient in $x$) describes the interactions in pristine EuO, while the linear and quadratic terms describe the overall changes in $T_C$ that arise due to barium inclusion.

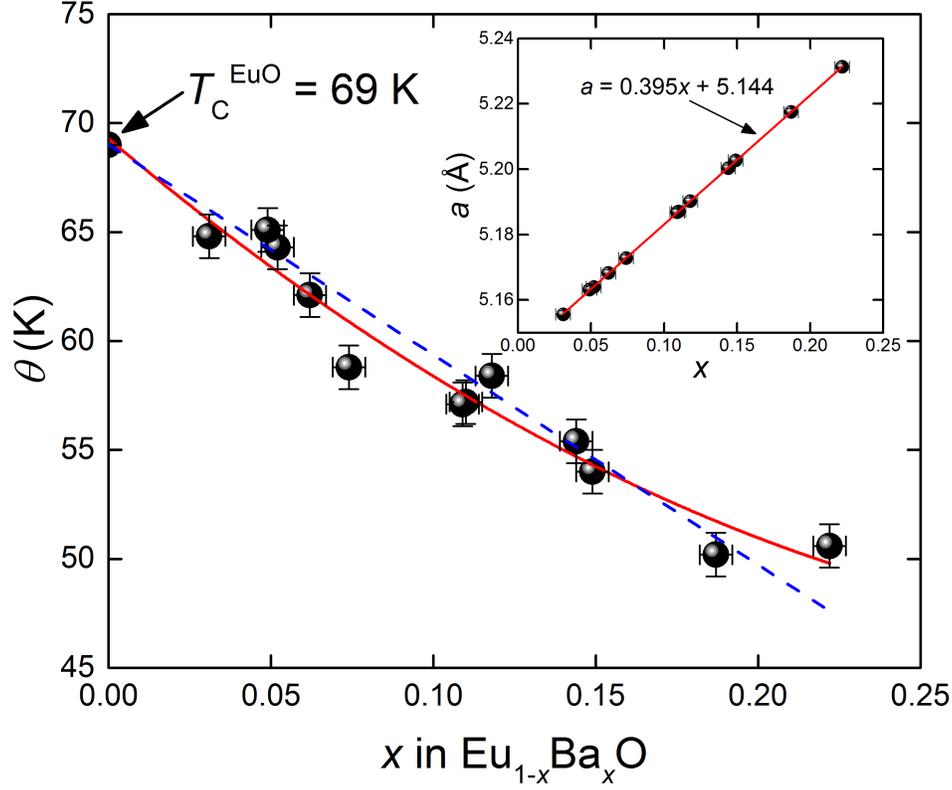

FIG. 5. Weiss constant as a function of barium content. The solid red line is the fit to the modified Heisenberg model in Eqn. 5, while the dashed blue line is a fit to the same model without a quadratic term. Inset: the lattice parameter $a$ depends linearly on the barium content $x$.

To exclude possible variations of $T_C$ due to variations of the Arrott criterion, we employ the measured Weiss temperatures $\theta$ instead and plot these as a function of $x$ in Fig. 5. In addition, the point at $x=0$ has been added with $T_C=69$ K to reflect the transition temperature of pristine EuO. The data including the point at $x=0$ is well explained by the model in Eqn. (5) (solid red line in Fig. 5), yielding the fit parameters $J_{\text{eff}}(0, a_0)/k_B \approx 6.59 \pm 0.09$ K, $\frac{\partial J_{\text{eff}}(0,a_0)}{\partial a}/k_B \approx -15.2 \pm 2.3$ K/Å, and $\frac{\partial^2 J_{\text{eff}}(0,a_0)}{\partial a^2}/k_B \approx 53 \pm 26$ K/Å$^2$. Excluding the point at $x=0$ and fitting *only* the experimental data yields a remarkably close fit (not shown in Fig. 5 as it is overlapping with the solid red line) with parameters 6.6±0.2 K, –16±4 K/Å, and 61±36 K/Å$^2$. It is clear that an extension of the model to third order will yield no new insight due to the large uncertainty in the second order coefficient, arising from the scatter in the data. Ignoring the quadratic term, the coefficient of the linear term provides the change in interaction strength due to lattice parameter changes. The value obtained in this study compares reasonably to the value of –17.04 K/Å obtained from neutron studies by Passell et al. [52], and acceptably to –7.44 K/Å estimated from pressure studies by Stevenson and Robinson [53].

It may also be noteworthy to compare the zeroth order term to reported EuO values in the literature, and particularly to $J_2$. By fixing $J_1$ to the widely accepted value $J_1/k_B=0.625$ K, we obtain via Eqn. (3'), $J_2/k_B \approx -0.15 \pm 0.02$ K, i.e., our study indicates that the NNN interactions are antiferromagnetic.

Neglecting the quadratic term and keeping the zeroth order term for pristine EuO, the linear fit is shown as a dashed blue line in Fig. 5, yielding the parameter $\frac{\partial J_{\text{eff}}(0,a_0)}{\partial a}/k_B \approx -11.6 \pm 0.5$ K/Å. This fit is a fair approximation, showing a reduction of $T_C$ with $x$ of approximately –96 K/Ba. It is therefore expected that a crystal of $Eu_{1-x}Ba_xO$ with $x \approx 0.72$ will have a ferromagnetic Curie temperature of 0 K, potentially producing a quantum phase transition in this system. This concentration value is above the expected percolation threshold for an fcc lattice and indicates that the long-range magnetic order is sustained with only about three of the twelve nearest neighbors carrying a spin.

**Conclusion**

Growth of $Eu_{1-x}Ba_xO$ single crystals has been achieved at temperatures below 1000°C using a barium-magnesium flux and europium sesquioxide as starting materials. A series of single crystals were grown from $x$=0.03 to 0.25, and synthesis of pristine EuO may be possible by further optimizing the growth parameters. All crystals grown from the alkaline metal flux are highly insulating, therefore ruling out an increase/change in the ferromagnetic Curie temperature due to free electrons. The observed changes in $T_C$ can therefore be ascribed exclusively to the dilution of the $Eu^{2+}$ magnetic lattice by non-magnetic $Ba^{2+}$ ions and the commensurate lattice expansion. The inclusion of the larger Ba ion produces a "negative" pressure on the lattice, therefore lowering the interaction strength. The trend in $T_C$ is explained by a modified Heisenberg model, and we observe a change in exchange interaction parameter with lattice parameter similar to reported values in the literature. We estimate that for $(Eu_{0.78}Ba_{0.22})O$, the lattice expansion is responsible for a decrease in $T_C$ of about 1.5 K, while the Ba substitution on the Eu lattice is responsible for a decrease of 18.2 K. While the effect of the lattice expansion is less than 10% of the overall reduction of $T_C$, it should not be neglected. Furthermore, engineering lattice strain will allow tuning $T_C$ over a similar temperature range.

We further infer that, to a first approximation, a crystal of $Eu_{1-x}Ba_xO$ with $x \approx 0.72$ should have a $T_C=0$ K, indicating the possibility of a quantum phase transition in this mixed Eu-Ba system. However, we surmise that an expanded study with larger Ba substitutions will reveal a larger lattice strain effect and a cubic dependency of $T_C$ on $x$, hinting at a quantum phase transition at a lower doping level than $x \approx 0.72$.

**Acknowledgment**

We gratefully acknowledge S. von Molnar for valuable discussions, and J. Neu for comments. This work is supported by the Department of Energy, Office of Basic Science, under contract DOE SC-0008832. Part of this work has been performed at the National High Magnetic Field Laboratory, which is supported by the National Science Foundation Cooperative Agreement DMR-1157490, the State of Florida, and the U.S. Department of Energy.